\documentstyle[epsf,epsfig,12pt]{article}

\newlength{\dinwidth}
\newlength{\dinmargin}
\newcommand{\ba}{\begin{array}}
\newcommand{\ea}{\end{array}}
\newcommand{\be}{\begin{equation}}
\newcommand{\ee}{\end{equation}}
\newcommand{\bea}{\begin{eqnarray}}
\newcommand{\eea}{\end{eqnarray}}
\newcommand{\beas}{\begin{eqnarray*}}
\newcommand{\eeas}{\end{eqnarray*}}

\def\to{\rightarrow}
\def\bee{\begin{eqnarray}}
\def\eee{\end{eqnarray}}
\def\be{\begin{equation}}
\def\ee{\end{equation}}

\def\laplace{{\kern1pt\vbox{\hrule height 1.2pt\hbox{\vrule width
1.2pt\hskip
  3pt\vbox{\vskip 6pt}\hskip 3pt\vrule width 0.6pt}\hrule height 0.6pt}
  \kern1pt}}
\def\scriptlap{{\kern1pt\vbox{\hrule height 0.8pt\hbox{\vrule width
0.8pt
  \hskip2pt\vbox{\vskip 4pt}\hskip 2pt\vrule width 0.4pt}\hrule height
0.4pt}
  \kern1pt}}

\def\roughly#1{\raise.3ex\hbox{$#1$\kern-.75em\lower1ex\hbox{$\sim$}}}

\begin{document}
\thispagestyle{empty}
\addtocounter{page}{-1}
\begin{flushright}
CLNS 97/1535\\
SNUTP 97-147\\
{\tt hep-ph/9712528}
\end{flushright}
\vspace*{1cm}
\centerline{\Large \bf Baryon Chiral Perturbation Theory}
\vskip0.4cm
\centerline{\Large \bf in}
\vskip0.4cm
\centerline{\Large  \bf Partially Quenched Large-$N_c$ QCD 
\footnote{
Work supported in part by NSF Grant, NSF-KOSEF
Bilateral Grant, KOSEF SRC-Program, Ministry of Education Grant BSRI
97-2418, SNU Research Fund, 
and the Korea Foundation for Advanced Studies Faculty Fellowship.}}
\vspace*{1.2cm}
\centerline{\large\bf Chi-Keung Chow${}^a$ and Soo-Jong Rey${}^b$}
\vspace*{0.6cm}
\centerline{\large\it Newman Laboratory for Nuclear Studies}
\vskip0.1cm
\centerline{\large\it Cornell University, Ithaca NY 14853 USA${}^a$}
\vskip0.45cm
\centerline{\large\it Physics Department, Seoul National University}
\vskip0.3cm
\centerline{\large\it Seoul 151-742 KOREA${}^b$}
\vspace*{0.8cm}
\centerline{\large\tt ckchow@mail.lns.cornell.edu,
sjrey@gravity.snu.ac.kr}
\vspace*{1.5cm}
\centerline{\large\bf abstract}
\vskip0.5cm
Baryons of lower spin in partially quenched large-$N_c$ QCD are studied 
with particular emphasis to interpolation between standard unquenched
and 
fully quenched limits. In large $N_c$ limit of partially quenched QCD,
we 
calculate $\Delta m$, the chiral one-loop correction to baryon masses.
We 
find that leading order 
contribution to $\Delta m$ is {\sl independent} of number of ghost
quarks introduced. For finite $N_c$, $\Delta m$ does satisfy the 
Bernard--Golterman's third theorem and has no infared quenching 
singularity except for fully quenched limit. At large $N_c$ limit,
however, 
we show that the third theorem is bypassed and non-trivial quenched
chiral 
corrections do arise. In unquenched limit, we also show that standard
chiral 
perturbation theory results are reproduced in which $\eta'$ loop
contributions 
are explicitly taken into account.  

\vspace*{2cm}

\setlength{\baselineskip}{18pt}
\setlength{\parskip}{12pt}
\newpage

\section{Introduction}
With impressive advance in numerical simulation of QCD on a lattice,
the goal of first-principle calculation for hadron spectrum as well as
other physical quantities is achievable in the near future. At present,
however, nearly all QCD simulations are based on various degrees of
approximation. 
Therefore, understanding error and deviation incurred by such
approximations to the calculated physical quantity is of prime
theoretical 
importance.
Among the approximations, adopted most frequently is the {\sl quenched 
approximation}~\cite{quenched1,quenched2}. By quenched approximation,
which 
has been introduced in order to accelerate generation of the gauge field
configurations, 
one drops out quark determinants and keeps only the valence quarks. 
While the approximation is well suited when quark masses are taken
sufficiently
heavy, extrapolation of quark masses to chiral limit might lead to 
potentially significant effects. 
In fact, Sharpe~\cite{sharpe1, sharpe2} and Bernard and
Golterman~\cite{bg1, 
bg2} have pointed out that the quenched approximation leads to sizable
errors in the chiral limit. If this is the case for a generic physical 
quantity, then one needs to understand better the errors incurred by the 
quenched approximation before a prediction or a conclusion is drawn. 

To investigate systematically an error introduced by the quenched
approximation
in the chiral limit, Sharpe~\cite{sharpe1, sharpe2} and Bernard and 
Golterman~\cite{bg1} have developed chiral perturbation theory for
quenched 
QCD.
This so-called quenched chiral perturbation theory (Q$\chi$PT) utilizes
the 
trick by Morel~\cite{morel} of introducing ghost quarks, which have the
same 
masses as the original quarks but carries opposite statistics. Thus, the 
chiral flavor symmetry is extended to graded, super-flavor symmetry and
it is 
this symmetry which organizes the Q$\chi$PT. Moreover,
diagrams with quark loops are cancelled by the same diagrams but with
ghost 
quark loops. 
Since these pioneering works, using the Q$\chi$PT, chiral behavior of
quenched 
QCD has been studied extensively for pseudo-scalar 
mesons~\cite{sharpe1, sharpe2, bg1}, baryons~\cite{labrenzsharpe}, 
vector and tensor mesons~\cite{boothfalk, chowrey2}, 
heavy mesons~\cite{booth, sharpezhang}, weak matrix elements 
$B_K$~\cite{sharpe2, goltermanleung}, baryon axial charge~\cite{kimkim}, 
heavy baryons~\cite{chiladze}, and pion scattering length~\cite{bg3}.
Numerous lattice data have been compared to results from Q$\chi$PT: see, 
reviews by Sharpe~\cite{sharperev}, Gottlieb~\cite{gottliebrev} and 
Okawa~\cite{okawarev} and references therein.

Through these extensive study, it has been concluded that main effect of 
quenched approximation is due to peculiarity of super-$\eta'$ meson
(direct 
counterpart of ``anomalous'' $\eta'$). This super-$\eta'$ meson has the
same 
mass as the other Goldstone bosons (same single-pole position of
propagators) 
and has to be retained in the Q$\chi$PT, unlike the standard $\chi$PT.
Peculiar to 
super-$\eta'$ 
is the extra double-pole term present in the propagator, which does not
permit a particle interpretation at all.
Most notably, it has been found that the double-pole term in the
super-$\eta'$ 
propagator leads to new infrared singular, non-analytic chiral
corrections that
have no counterpart in the full QCD. In order to understand better such 
pathological behavior, it should be desirable to be able to interpolate, 
if possible at all, QCD between unquenched and fully quenched limits and 
examine changes of
physics in the chiral limit. Indeed, with such motivation, Bernard and 
Golterman~\cite{bg2} have developed partially quenched QCD, in which
only a subset of quarks are paired with ghost quarks~\footnote{As
Bernard and
Golterman point out, partially quenched QCD arises also quite naturally
for description of lattice QCD with staggered fermions or of lattice QCD
with Wilson valence quarks and staggered sea-quarks.}. 
By formulating partially quenched chiral perturbation theory
(PQ$\chi$PT), 
Bernard and Golterman have shown that change of chiral behavior for 
Goldstone mesons can be indeed understood from interpolation between
standard 
QCD and fully quenched QCD.

In this paper, we study chiral dynamics of large $N_c$ 
{\sl baryons in partially 
quenched QCD}. Previously, in the fully quenched QCD, baryons have been 
studied by Labrenz and Sharpe~\cite{labrenzsharpe}. 
They have studied quenching effect to the behavior of baryon mass
spectrum 
in the chiral limit and have observed, most notably, that pattern of the 
correction is substantially different from that in the unquenched case. 
The leading order chiral correction in Q$\chi$PT scales like
$m_q^{1/2}$, 
which is more singular than the standard $m_q^{3/2}$ $\chi$PT
correction.   
Prompted by such interesting results, 
in this paper, by formulating partially quenched chiral perturbation
theory for baryons, we examine change in 
chiral behavior of baryon mass spectrum as the QCD interpolates between
standard and fully quenched limits. 
In our investigation, we also invoke large $N_c$ limit and associated 
planar symmetry. Combined application of chiral perturbation theory and 
large $N_c$ expansion  
are expected to constrain the low-energy interactions of baryons with
the
Goldstone bosons more effectively than either method alone.
We have found that the large $N_c$ limit affects the chiral dynamics
in two interesting ways. First, the positions (on the complex plane) of 
the poles of the $\eta'$ propagator depend explicitly on the value of
$N_c$.  
In PQ$\chi$PT, the $\eta'$ meson propagator has two simple poles: pion
pole
at $p^2=m^2$ and a shifted pole at $p^2=m^2+M_0^2$.  
The ``pole shift'' $M_0^2$ originates from the hairpin diagram, which is 
$1/N_c$ suppressed.  As $N_c\to \infty$, the shifted pole will merge to 
the pion pole. As we will show below, there arises an interesting
cancellation 
between contributions from each of these two poles in the $\eta'$
propagator.
Second, the size of $N_c$ also controls the form of axial current
couplings 
in the baryon sector.  
The $\eta'$ coupling to {\it a single quark line\/} inside a baryon is
related to 
its pion counterpart, as $\pi$ and $\eta'$ are related by the ``planar 
symmetry'', which is exact in the large $N_c$ limit.  
This symmetry, however, is {\it not\/} manifest at the hadron level. 
While pion coupling to different quark lines interfere constructively,
the $\eta'$ coupling has huge cancellation.  
As a result, chiral loops involving $\pi$ and $\eta'$ appear at
different order in large $N_c$ expansion.  

This paper is organized as follows. In section 2, after recapitulation
of
baryons in the large-$N_c$ limit we formulate partially quenched chiral 
perturbation theory for large $N_c$ baryons. We pay particular attention
to the mass corrections of nucleons and $\Delta$ in the large $N_c$ and
chiral limits. 
In section 3, we calculate infrared singular, non-analytic chiral 
corrections to the baryon masses. Moreover, we provide anatomy of chiral 
one-loop corrections and identify contributions of $\eta'$ at leading
order 
in $1/ N_c$ expansion. In section 4, we investigate the effect of next
leading order corrections in $1/N_c$ expansions, followed by some
discussions. 

\section{Baryons in Partially Quenched Large-$N_c$ QCD}

\subsection{Baryon Dynamics in Large $N_c$ QCD}
Large $N_c$ meson dynamics has been first studied in Ref.~\cite{T}.   
Baryon dynamics in large-$N_c$ QCD has been studied originally in 
Ref.~\cite{witten} and, more recently, in
Refs.~\cite{DJM1,CGO,LM,DJM2,J}. 
We recapitulate essential aspects of their results that will become 
relevant for baryons in  partially quenched QCD (PQ-QCD). For
simplicity, 
we will mainly focus on the case with $n = 2$ ($u$ and $d$).
Since the $N_c$ quarks inside a baryon should form a color singlet,
which is 
completely antisymmetric, the spin-flavor part of the wave function must 
be completely symmetric.  
As a result, the lowest lying baryons have ${\bf I}={\bf J}={1\over2}$, 
${3\over2}, \dots$, which are usually identified as the observed states
N, 
$\Delta, \dots$ 
\footnote{See Ref.~\cite{G} for detailed discussion on validity of such 
an identification.}.  

The interaction of baryons with Goldstone bosons are simplified in the 
large $N_c$ limit as the dynamical symmetries are enlarged both in the 
meson and the baryon sector.  
In the Goldstone boson sector, there arises the {\tt planar symmetry}
(also 
called ``nonet symmetry'' in the literature), decreeing that the $\eta'$ 
meson should be combined with the other Goldstone bosons into a 
$n\times n$ representation of U($n$) flavor symmetry group, where $n$ is
the number of light flavors \cite{T,witten}.  
More explicitly, for $n = 2$, Goldstone boson fields are represented by
the $2 \times 2$ matrix: 
\begin{equation}
\phi = \pi^a {\bf T}^a + \eta' {{\bf 1} \over \sqrt 2} 
= \phi^\alpha {\bf T}^\alpha = 
\pmatrix{{\eta'\over\sqrt{2}}+{\pi^0\over\sqrt{2}}&\pi^+\cr
\pi^-&{\eta'\over\sqrt{2}}-{\pi^0\over\sqrt{2}}\cr}, 
\end{equation}
where ${\bf T}^a$ ($a=1$, 2, 3) are the SU(2) generators, {\bf 1} is the 
identity, and ${\bf T}^\alpha$ ($\alpha=0$, 1, 2, 3) are the U(2)
generators, 
{\it i.e.}, $\{{\bf T}^\alpha\} = \{{\bf T}^a\} \cup \{{\bf 1}\}$.  
As we will see below, the $N_c \rightarrow \infty$ planar symmetry
implies 
that the same set of coupling constants will control both the 
$\pi$ and the $\eta'$ interactions. 

In the large $N_c$ limit, the Goldstone boson kinetic term is 
\begin{equation}
{\cal L}_0={f^2\over8}{\rm
Tr}\left(\partial^\mu\sigma\,\partial_\mu\sigma 
\right) + \dots, 
\end{equation}
where 
\begin{equation}
\sigma \equiv \exp(2i\phi/f) ,  
\end{equation}
and ellipses denote higher derivative interactions that are irrelevant
for 
low-energy dynamics.   
The degeneracy between $\pi$ and $\eta'$ persist even if one 
endows a small mass to the light quarks and breaks the chiral symmetry.
The quark mass leads to a non-zero mass $m_\pi^2 =m_{\eta'}^2 = m^2$.  
The U(1)$_A$ anomaly, which breaks the planar symmetry and renders a 
heavy mass to $\eta'$, shows up as an $1/N_c$ suppressed correction:  
\begin{equation}
{\cal L}_1=\textstyle{1\over2} \bigg(m_0^2 ({\rm Tr} \,\phi)^2 
+ A_0(\partial_\mu {\rm Tr} \,\phi)^2\bigg) 
={n\over2}\left( \, m_0^2\,\eta'^2+A_0\,\partial_\mu\eta'\,
\partial^\mu\eta' \, \right), 
\end{equation}
where $n$ is the number of light flavors, and $A_0$ and $m_0^2$ are of
order 
${\cal O}(N_c^{-1})$.
In the standard, unquenched $\chi$PT, the effect of ${\cal L}_1$ can be 
resummed:
\begin{eqnarray}
&{}& 
\sum_{k=0}^\infty {1 \over p^2 - m^2} \Big( \left( n \, A_0 p^2 + n \,
m_0^2 
\right) { 1 \over p^2 - m^2} \Big)^k
\nonumber \\
&=& 
{1\over 1-n\,A_0}{1\over p^2 - (m^2+n\,m_0^2/(1-n\,A_0))}. 
\label{ep}
\end{eqnarray}
One finds that new position of single pole at $m_{\eta'}^2 = m^2 +
nm_0^2/(1-nA_0)$.  \footnote{In the literature, $m_0^2$ and $A_0$ are
often 
written as $\mu^2/3$ and $\alpha/3$ so the for $n=3$, $m_{\eta'}^2 = m^2
+ \mu^2/(1-\alpha)$.  
In this paper, however, we naturally opted for the notation without the 
factors of $1/3$ as we are mainly working with just two light flavors.} 
Thus, in the chiral limit, $m_{\eta'}^2 \sim {\cal O}(N_c^{-1})$, as
expected.  

On the other hand, the baryon sector exhibits an SU($2n$) {\tt
spin-flavor 
symmetry}~\footnote{Our description of the spin-flavor symmetry for 
large $N_c$ baryons follow closely Ref.~\cite{CGO,DJM2,G}.  
Part of the reason is that the formalism in Ref.~\cite{DJM1}, which has
been
widely employed in the literature, assumes unitarity, hence, cannot be 
directly applicable to non-unitary, (partially) quenched QCD we study
presently. However, the physical predictions of any of these formalisms 
should be identical.} .
For simplicity, we will restrict ourselves with 2 light flavors for
foregoing
discussions, in which case the spin-flavor SU(4) symmetry will be
generated 
by: 
\begin{equation}
{\bf J}^i = \sum_{k=1}^{N_c} q_k^\dag ({\bf S}^i \otimes {\bf 1}) q_k,
\quad
{\bf I}^a = \sum_{k=1}^{N_c} q_k^\dag ({\bf 1} \otimes {\bf T}^a) q_k,
\quad
{\bf G}^{ia} = \sum_{k=1}^{N_c} q_k^\dag ({\bf S}^i \otimes {\bf T}^a)
q_k \, .  
\end{equation}
Here, the spin and isospin operators on individual quark lines are
denoted by 
${\bf S}^i$ ($i=x$, $y$ and $z$ are three spacelike directions
perpendicular 
to the baryon velocity) and ${\bf T}^a$ respectively, and {\bf 1} is the 
identity operator in spin and isospin spaces.  
Note that there are 3 {\bf J}'s, 3 {\bf I}'s and 9 {\bf G}'s, making up
the correct number of generators Eq. (15) for the spin-flavor SU(4)
algebra.  
In what follows, we will adopt a useful notation
\begin{equation}
{\bf G}^{i\alpha} = \sum_{k=1}^{N_c} q_k^\dag ({\bf S}^i \otimes 
{\bf T}^\alpha) q_k, 
\end{equation}
which is equal to ${\bf J}^i$ and ${\bf G}^{ia}$ for zero and non-zero 
$\alpha$ respectively.  

The axial current couplings of large $N_c$ baryons are simplified by the 
fact that multi-quark operators are suppressed by powers of $1/N_c$.  
In the leading order of the large $N_c$ expansion, the axial currents
couple 
through single-quark operators.  
Hence, interaction of baryons to Goldstone bosons can be expressed 
as~\cite{CGO,J}: 
\begin{eqnarray}
{\cal L}_{\rm quark} &=& i{g\over f} \partial^i \phi^\alpha G^{i\alpha} 
\nonumber\\&=& i{g\over f} \left(
\partial^i \pi^a \,\sum_{k=1}^{N_c} q_k^\dag\,({\bf S}^i\otimes{\bf
T}^a)\,q_k 
+\partial^i \eta'\,\sum_{k=1}^{N_c} q_k^\dag\,({\bf S}^i\otimes {\bf
1})\,q_k 
\right) + \cdots , 
\label{I}
\end{eqnarray}
where $f = f_\pi = f_{\eta'}$ denotes Goldstone boson decay constant, 
and ellipses denote higher-order interactions involving more than one 
Goldstone bosons.  
Note that the $\pi$ and $\eta'$ couplings to baryons
are taken by the same coupling constant $g$, as decreed by planar  
symmetry in the large-$N_c$ limit.

One can read off from Eq.~(\ref{I}) the meson-baryon-baryon couplings by 
summing up contributions from each individual quark lines. 
The $\eta'$ meson couples to individual quark through 
${\bf S}^i \otimes {\bf 1}$. Hence, $\eta'$ meson coupling to 
a {\it baryon} obtained by summing over all constituent quarks is given
by
spin ${\bf J}^i$ of the baryon. That is,  
\begin{equation}
{\cal L}_{\eta'BB} \quad = \quad i{g \over f} \left( \partial^i \eta'
\right) 
\,\sum_{k=1}^{N_c} q_k^\dag \, ({\bf S}^i \otimes {\bf 1}) \, 
q_k \quad = \quad i{ g \over f} \left( \partial^i \eta' \right) \, 
{\bf J}^i \quad .
\end{equation}
Since we are interested only in states with ${\bf I}={\bf J}\sim {\cal
O} 
(N_c^0)$, we will find that the $\eta'BB$ coupling is of order ${\cal O} 
(N_c^{-1/2})$ (as $f\sim {\cal O}(N_c^{1/2})$).  
Note that ${\cal L}_{\eta'BB}$ contains sum over $N_c$ quark terms, each 
of them being of order $1/f = {\cal O}(N^{-1/2})$. Hence, one would
naively 
suspect that the sum to be of order $(N_c/f) \approx {\cal
O}(N_c^{1/2})$.  
This reasoning is fallacious, however, as for lower spin states, there
exists huge cancellations among the individual quark spins.  

One the other hand, pion coupling is proportional to 
${\bf S}^i \otimes {\bf T}^a$, and acts coherently over the individual 
quarks~\cite{CGO,LM,DJM2,J,G}. That this is so can be seen clearly 
using the operator identity~\cite{DJM2,G}: 
\begin{eqnarray}
&& \sum_i \left( \sum_{k=1}^{N_c} 
q_k^\dag \, ({\bf S}^i \otimes {\bf 1}) \, q_k \right)^2 
+ 
\sum_a \left( \sum_{k=1}^{N_c} 
q_k^\dag \, ({\bf 1} \otimes {\bf T}^a) \, q_k \right)^2 
+
4\sum_{i,a} \left( \sum_{k=1}^{N_c} 
q_k^\dag \, ({\bf S}^i \otimes {\bf T}^a) \, q_k \right)^2 
\nonumber \\
\nonumber \\
&& \qquad = {\bf J}^2 + {\bf I}^2 + 4 {\bf G}^2 
= (\textstyle{3\over4}N_c^2 + 3N_c) \, {\bf 1},  
\end{eqnarray}
which is nothing but the Casimir identity for the spin-flavor SU(4).
Since both ${\bf J}^2$ and ${\bf I}^2$ are of order ${\cal O}(N_c^0)$,
the 
operator ${\bf G}^{ia}=\sum_{k=1}^{N_c}q^\dag_k\,({\bf S}^i\otimes{\bf
T}^a)\,
q_k$ must be of order ${\cal O}(N_c)$ so that the sum is of order 
${\cal O}(N_c^2)$.  
As a result, $\eta'$ coupling is one order in $1/N_c$ smaller than the
pion counterpart, and we can see immediately that $\eta'$ loops will be
suppressed by $1/N_c^2$ with respect to pion loops.  
Since all infared quenched singularities arise from $\eta'$ loops, we
expect 
these infared quenched singularities are subleading in orders of
$1/N_c$.  
Below, we will show explicitly that this is indeed the case.  

\subsection{PQ$\chi$PT for Large-$N_c$ Baryons} 
Now we are ready to (partially) quench the theory.  
Following the example of Ref.~\cite{bg2}, one includes in the theory, 
in addition to the $n$ quarks, $k$ ghosts (spin-{$1\over2$} objects with 
bosonic statistics), where $k\le n$.  
The $n\times n$ Goldstone boson matrix $\phi$ gets enlarged to an
$(n+k)\times 
(n+k)$ supermatrix $\Phi$, 
\begin{equation}
\Phi \equiv \pmatrix {\phi & \chi^\dag\cr \chi & {\tilde \phi}\cr}, 
\end{equation}
where $\chi$ is the $n\times k$ matrix of quark-antighost fermionic
``mesons'',
its charge conjugate $\chi^\dag$ the $k\times n$ matrix of
ghost-antiquark 
``mesons'', and $\tilde \phi$ the $k\times k$ matrix of ghost-antighost 
bosonic mesons.  
The field $\Phi$ transforms as a $(n+k)\times\overline{(n+k)}$ 
representation of the enlarged 
symmetry algebra U($n+k$) (or more exactly, the U($n|k$) graded
algebra).  
We will denote the generators of this U($n+k$) by ${\bf T}^A$ and 
\begin{equation}
\Phi = \Phi^A {\bf T}^A.  
\end{equation}
Note that $\{{\bf T}^a\}\subset\{{\bf T}^\alpha\}\subset\{{\bf T}^A\}$.  

The lagrangians ${\cal L}_0$ and ${\cal L}_1$ are generalized to
\cite{bg2} 
\begin{equation}
{\cal L}_0={f^2\over8}{\rm Str}(\partial^\mu\Sigma\partial_\mu\Sigma) 
+ \dots, 
\end{equation}
\begin{equation}
{\cal L}_1={1\over2} \bigg(m_0^2 ({\rm Str} \,\Phi)^2 
+ A_0(\partial_\mu {\rm Str} \,\Phi)^2\bigg) 
\end{equation}
where 
\begin{equation}
\Sigma \equiv \exp(2i\Phi/f),  
\end{equation}
and ``Str'' denotes the supertrace, the sum of the first $n$ entries on
the diagonal subtracted by the sum of the last $k$ entries.  
The propagator for flavored states in $\Phi$ has a simple pole at $p^2 = 
m^2$, while for the flavor-neutral states (states on the diagonal of
$\Phi$) 
the propagator is \cite{bg2}
\begin{equation}
G_{ij} = \Big[ {\delta_{ij} \epsilon_i \over p^2 - m^2} + {1\over\Delta
n} 
\big({1 \over (1 + \Delta n\, A_0)\, p^2 - (m^2 + \Delta n\, m_0^2)} - 
{1 \over p^2 - m^2}\big)\Big].
\label{pro}
\end{equation}
The grading index $\epsilon_i$ is such that
$\epsilon (q_i) = + 1, \epsilon ({\tilde q}_j) = -1$.  
Note that the second term (which proportional to $1/\Delta n$) mixes 
quark-antiquark meson with ghost-antighost mesons.  
Both the first and the second terms have a simple pole at $p^2 =
m^2$, but 
the second term has an additional pole, which we will call the ``shifted 
pole''. 
Also note that, as $n \rightarrow k$, the shifted pole moves back to the
pion pole.
The propagator will then have a double pole at $m^2$, which is a
well-known result in Q$\chi$PT \cite{bg1}.  

The partial quenching of the baryon chiral lagrangian is
straightforward.  
The more general PQ$\chi$PT lagrangian in the large $N_c$ limit is 
\begin{equation}
{\cal L} = i{1\over f} \Big(g\, \left(\partial^i \Phi^A \right) 
\,{\bf G}^{iA} + h\, \left(\partial^i {\rm Str} \Phi \right)\, 
{\bf J}_\epsilon^i \, \Big).   
\end{equation}
The first term is a simple generalization of the unquenched lagrangian,
with 
\begin{equation}
{\bf G}^{iA} = \sum_{k=1}^{N_c} q_k^\dag ({\bf S}^i \otimes {\bf T}^A)
q_k, 
\end{equation}
and $g$ an undetermined coupling constant.  
The second term describes the ``hairpin coupling'' which couples 
flavor-neutral states in $\Phi$ to the baryon operator ${\bf
J}_\epsilon$, 
defined by
\begin{equation}
{\bf J}_\epsilon^i = \sum_{k=1}^{N_c} q_k^\dag ({\bf S}^i \otimes 
{\bf 1}_\epsilon) q_k, 
\end{equation}
where ${\bf 1}_\epsilon$ is the identity of U($n|k$), with $n$ 1's and
$k$ 
$-1$'s on the diagonal.  
In this paper, however, we will set the hairpin coupling constant $h$ to 
zero to simplify the physics.  
A discussion of the possible effects of a non-zero $h$ will be given 
wherever appropriate in later sections.

\section{Non-analytic Chiral Correction to Baryon Masses}

\subsection{Chiral One-loop Correction}
We are now ready to calculate the chiral one-loop correction to the
baryon masses in PQ$\chi$PT.   
While our theory has $n$ quarks and $k$ ghosts, all assumed to be
degenerate, 
we will focus on baryons with just $u$ or $d$ quarks, though the non
$(u,d)$ 
quarks (which will be collectively referred as strange quarks) may
appear in 
loops.  
All the diagrams involved have the form as shown in Fig.~1 and yield the
same one-loop Feynman integral: 
\begin{equation}
{\cal I}_1(m^2) \equiv -{1\over12\pi f^2} \, m^3  
\end{equation}
as those in the standard $\chi$PT.
We will restrict, in this paper, to the choice of hairpin coupling
$h=0$, 
the ghost-antighost meson 
does not contribute, and the only relevant coupling is the Lagrangian 
Eq.~(\ref{I}).  

\begin{figure}
\vspace{1cm}
\hspace{3cm}
\epsfig{file=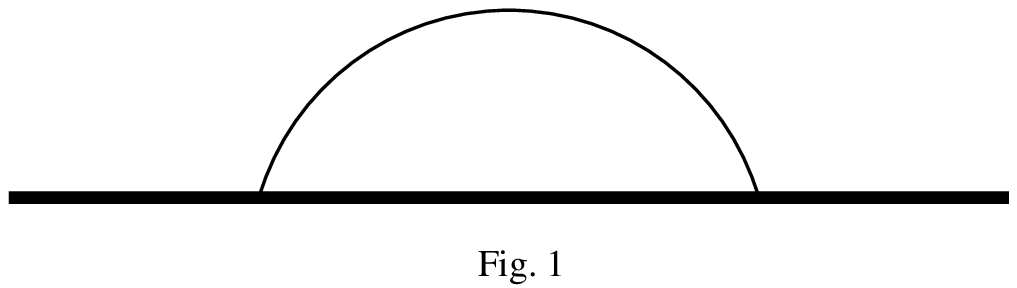, clip=}
\vspace{0.5cm}
\caption{Chiral one-loop diagrams to baryon two-point function.
Horizontal line denotes a baryon and upper internal line
denotes  Goldstone meson multiplets.}
\vspace{1cm}
\end{figure}

Depending on whether the Goldstone boson couples to the same quark line 
at the two vertices or not, there are two different types of
contribution to
the 
mass correction $\Delta m$.  

\begin{figure}
\vspace{1cm}
\hspace{3cm}
\epsfig{file=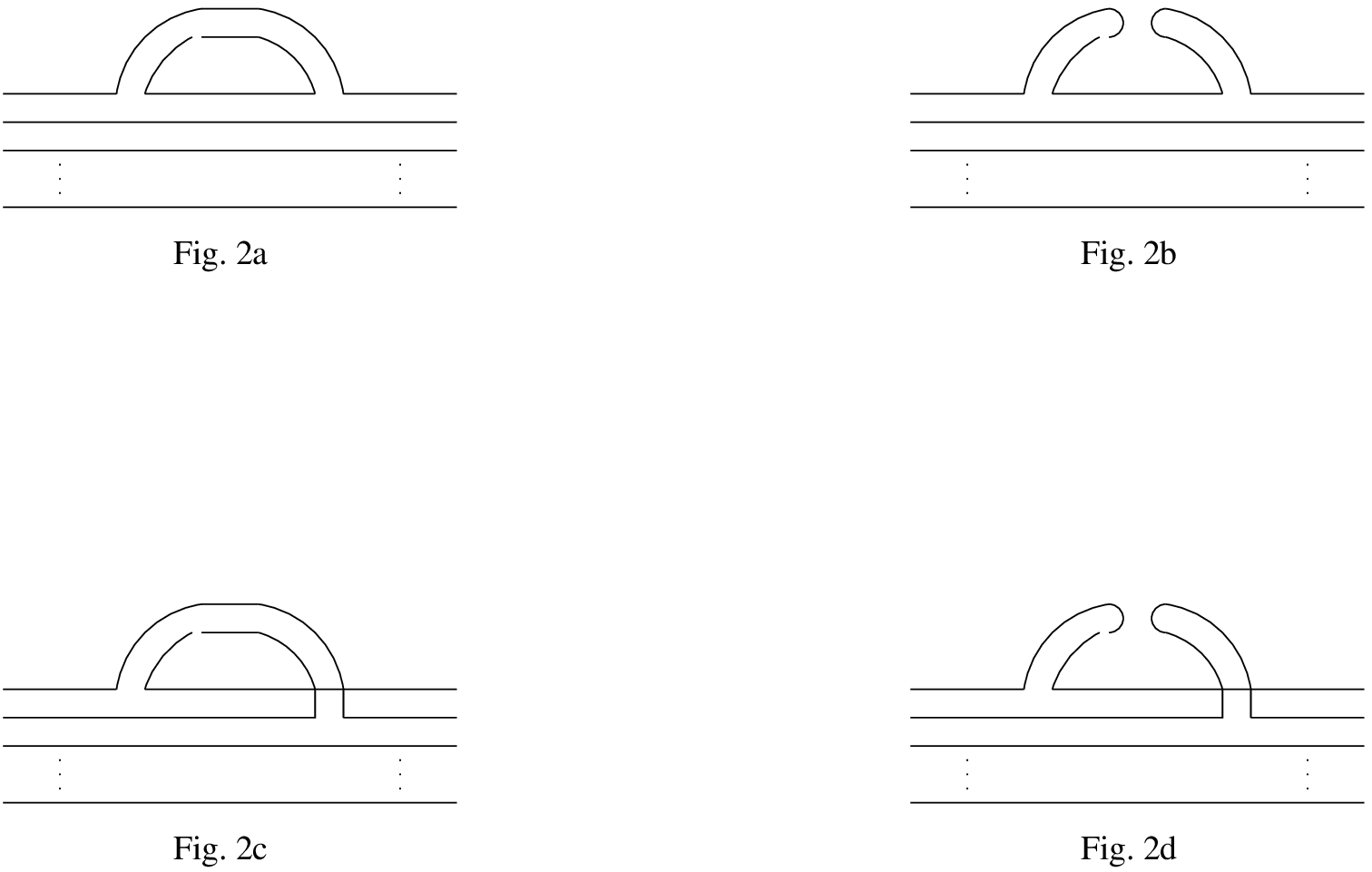, clip=}
\vspace{0.5cm}
\caption{Quark line diagarm for chiral one-loop correction to baryon 
two-point function. (a) Goldstone boson coupling to the same quark line 
, (b) hairpin propagator to the same quark line, (c) Goldstone boson
coupling to the different quark lines, (d) hairpin propagator to the 
different quark lines.}
\vspace{1cm}
\end{figure}

{\tt (i) 1 quark line chiral correction}: this class of
corrections arises when in
which the Goldstone meson couples to the same quark line at both
vertices.  
The quark level diagrams are Fig.~2a and 2b.  
Notice that Fig.~2b coupling is possible only when the intermediate 
Goldstone meson state is $\eta'$.  
The contribution of this class is given by 
\begin{eqnarray}
\Delta m = g^2 \, \, \, \sum_{k=1}^{N_c} \!\!\! &\Big[& \,  
{\cal I}_1 (m^2) \cdot 
\left( q_k^\dag \, ({\bf S}^i \otimes {\bf T}^A) \, q_k \right)^2 
\nonumber \\
&+& {2 \over \Delta n} \cdot \big({\cal I}_1(M^2) - {\cal I}_1(m^2)\big)
\cdot 
\left(q_k^\dag \, ({\bf S}^i \otimes {\bf 1}) \, q_k \right)^2 \Big].  
\end{eqnarray}
The first line comes from the chiral loops with flavored Goldstone
mesons, 
and the flavor-neutral mesons propagating with the first term in
propagator Eq.~(\ref{pro}).  
Note that the sum over U($n|k$) flavor generators ${\bf T}^A$ originates 
from the internal loop in Fig.~2a, which can be either a quark or a
ghost.  
The second term line originates from the second term in Eq.~(\ref{pro}), 
with 
\begin{equation}
{\cal I}_1(M^2) = {1\over 1+ \Delta n \cdot A_0} \cdot {\cal I}_1
\left({m^2 + \Delta n \cdot m_0^2 \over 1 + \Delta n \cdot A_0} \right). 
\end{equation} 

{\tt (ii) 2 quark line chiral correction}: this is the class of
corrections in
which the Goldstone meson couples to different quark lines at each
vertices.  
The quark level diagrams are depicted in Fig.~2c and 2d. The
contribution is 
given by 
\begin{eqnarray}
\Delta m = g^2 \sum_{k,\ell=1}^{N_c}\!\!{}' \Big[ 
\!\! && \!\!\! {\cal I}_1 (m^2) \cdot 
\left( q_k^\dag \, ({\bf S}^i \otimes {\bf T}^\alpha) \, q_k \right) 
\cdot\left( q_\ell^\dag \, ({\bf S}^i \otimes {\bf T}^\alpha)\, q_\ell
\right) 
\nonumber \\&+& 
\left( {2 \over \Delta n}\right)\cdot\big({\cal I}_1(M^2)-{\cal
I}_1(m^2)\big)
\cdot\left(q_k^\dag \, ({\bf S}^i \otimes {\bf 1}) \, q_k \right) \cdot
\left( q_\ell^\dag \, ({\bf S}^i \otimes {\bf 1}) \, q_\ell \right)
\Big].  
\end{eqnarray}
The prime on the summation sign denotes the condition $k\neq\ell$.  
Interpretation of different terms is identical with that of 
{\tt (i)}.  
Note that in the first line, even though the sum should be in principle
over 
all possible Goldstone bosons in $\Phi$, there is actually no internal 
quark/ghost loops in Fig.~2c, hence, the ghosts do not contribute.  
As a result, we can just sum over intermediate states in $\phi$ instead, 
leading to a sum over ${\bf T}^\alpha$ instead of ${\bf T}^A$.  

Now it is time to perform the sums over the quark lines.  
First, we have terms involving the SU(2) generators ${\bf T}^a$, 
{\it i.e.}, the pion loop contributions.  
\begin{eqnarray}
\Delta m_{\pi {\rm -loop}} &=& g^2  {\cal I}_1 (m^2) \cdot
\sum_{k,\ell=1}^{N_c}
\left( q_k^\dag \, ({\bf S}^i \otimes {\bf T}^a) \, q_k \right)
\left( q_\ell^\dag \, ({\bf S}^i \otimes {\bf T}^a) \, q_\ell \right) 
\nonumber \\
&=& g^2 {\cal I}_1 (m^2) \left({3\over8}N_c^2 +
{3\over2}N_c - {1\over2}{\bf J}^2 \right), 
\end{eqnarray}
where we have used the identity \cite{DJM2}
\begin{equation}
\sum_{k,l=1}^{N_c} 
\left(q_k^\dag \, ({\bf S}^i \otimes {\bf T}^a) \, q_k \right)
\left(q_l^\dag \, ({\bf S}^i \otimes {\bf T}^a) \, q_l \right)
=\left( \sum_{k=1}^{N_c} 
q_k^\dag \, ({\bf S}^i \otimes {\bf T}^a) \, q_k  \right)^2 
= \textstyle{3\over8}N_c^2 + \textstyle{3\over2}N_c -
\textstyle{1\over2} {\bf J}^2.  
\end{equation}

Second, we have terms involving the flavor identity {\bf 1}, {\it i.e.},
the $\eta'$ loop contributions.  
\begin{eqnarray}
\Delta m_{\eta' {\rm -loop}} &=& g^2 \left[{\cal I}_1 (m^2) +
{2\over\Delta n} 
\left({\cal I}_1 (M^2) - {\cal I}_1 (m^2)\right) \right] \cdot
\sum_{k,\ell=1}^{N_c} \left( q_k^\dag \,({\bf S}^i \otimes {\bf
1})\,q_k\right)
\left( q_\ell^\dag \, ({\bf S}^i \otimes {\bf 1}) \, q_\ell \right)
\nonumber\\
&=& g^2 
\left[{\cal I}_1 (m^2) + {2\over\Delta n} \left({\cal I}_1 (M^2) - 
{\cal I}_1 (m^2) \right)\right] {\bf J}^2, 
\end{eqnarray}
where we have used another identity, again from \cite{DJM2}
\begin{equation}
\sum_{k,l=1}^{N_c}
\left( q_k^\dag \, ({\bf S}^i \otimes {\bf 1}) \, q_k \right)
\left( q_l^\dag \, ({\bf S}^i \otimes {\bf 1}) \, q_l \right) 
= \left(\sum_{k=1}^{N_c}
q_k^\dag \, ({\bf S}^i \otimes {\bf 1}) \, q_k \right)^2={\bf J}^2.
\end{equation}

Finally, there are terms involving ${\bf T}^{\bar A}$, those ${\bf T}^A$
which do not belongs to the set of ${\bf T}^\alpha$.  
Physically these come from Fig.~2a, with a ghost or a strange quark
running 
around the internal loop.  
There are $n-2$ strange quarks and $k$ ghosts, so the contribution is 
\begin{equation}
\Delta m_{K/\chi {\rm -loop}} = (\Delta n - 2) \Delta m_{K {\rm -loop\;
with\;
1\; strange\; quark}}.  
\end{equation}
The combinatorics factor of the latter has been calculated, in 
Ref.~\cite{DJM1,DJM2} to be ${3\over4}N_c$. 
Hence, 
\begin{equation}
\Delta m_{K/\chi {\rm -loop}} = g^2 {\cal I}_1 (m^2) (\Delta n - 2) 
\textstyle{3\over4} N_c. 
\end{equation}

Combining all the contributions, we have chiral correction to baryons in 
partially quenched large $N_c$ QCD:  
\begin{equation}
\Delta m = + \,\, g^2 \,\, \Big[ \quad \left(
\textstyle{3\over8} N_c^2 +\textstyle{3\over4} \Delta n
N_c + {1\over2} {\bf J}^2 \right) \cdot {\cal I}_1(m^2) + 2 \cdot 
{1\over\Delta n} \cdot {\bf J}^2 \cdot \left(
{\cal I}_1(M^2) - {\cal I}_1 (m^2) \right) \quad \Big].
\label{m}
\end{equation}
Note that the final result depends only on $\Delta n$, which counts the 
difference in the number of physical quarks $n$ and the number of ghost 
quarks $k$, but not on $n$ and $k$ seperately.  Physically this reflects
the fact that physical quantities should not be changed upon
introduction 
of an extra set of degenerate quark and ghost quark pair, as the 
contribution should cancel out completely.  
To sum up, Eq.~(\ref{m}) is the mass correction to large $N_c$ baryons
in
PQ$\chi$PT.  
Note that this single expression works for all states with spin {\bf J}
as long as ${\bf J}\ll N_c$.  

Before we move on, we note that our result can be expressed in terms of
just 
${\cal I}_1$, the same functional form as standard $\chi$PT
results.  
The ``new chiral singularities'' or ``quenched infared divergences''
in Q$\chi$PT do not appear except the limit $\Delta n\to 0$.  
This is in agreement with the Bernard--Golterman's third theorem
\cite{bg2}, 
which states that {\sl quenched infared divergences appear if and only
if 
one or more of the valence quarks are fully quenched}.  Since the theory
we are considering is only partially quenched, we do not see the new
chiral 
singularities.  

\subsection{The Large-$N_c$ Decomposition of Chiral Corrections}
So far, we have analyzed the chiral one-loop correction to baryon mass 
spectrum without explicit reference to large-$N_c$ counting. 
In this section, we will decompose the chiral one-loop correction
explicitly as  a function of three distinct parameters: chiral symmetry 
breaking parameter $m_\pi$, planar symmetry parameter $1/N_c$ and
quenching parameter $\Delta n$.  

The chiral mass correction Eq.~(\ref{m}) contains three terms with
different 
$N_c$ and $\Delta n$ dependences:  
\begin{equation}
\Delta m = \Delta m_{(+1,0)} + \Delta m_{(0,+1)} + \Delta m_{(-1,0)} + 
\Delta m_{(-1,-1)}, 
\end{equation}
with 
\begin{eqnarray}
\Delta m_{(+1,0)} &= \textstyle{3\over8} \cdot g^2 \cdot N_c^2 \cdot 
{\cal I}_1(m^2) \hskip1.5cm &\sim {\cal O}\left(N_c^1, \Delta n^0
\right), 
\\
\Delta m_{(0,+1)} &= \textstyle{3\over4} \cdot g^2 \cdot \Delta n \cdot
N_c 
\cdot {\cal I}_1(m^2) \hskip1cm &\sim {\cal O}\left(N_c^0, \Delta n^1
\right),  
\\
\Delta m_{(-1,0)} &= \textstyle{1\over2} \cdot g^2 \, {\bf J}^2 \cdot
{\cal I}_1(m^2) \hskip1.7cm &\sim {\cal O} \left(N_c^{-1}, \Delta n^0
\right) 
\\
\Delta m_{(-1,-1)} &=  g^2 \, {\bf J}^2 \cdot {2\over\Delta n} \cdot 
\left({\cal I}_1(M^2) - {\cal I}_1(m^2) \right) 
&\sim {\cal O} \left(N_c^{-1}, \Delta n^{-1} \right).  
\end{eqnarray}
where we have pulled out implicit $1/N_c$ dependence through $1/f^2$ in
our definition of ${\cal I}_1$.
We immediately observe that the terms that depend on ${\bf J}^2$, which 
contribute to $\Delta$--N mass splitting, appear only at order
${\cal O}(N_c^{-1})$. 
This is in accordance with the large $N_c$ counting given in 
Ref.~\cite{DJM1,CGO,LM,DJM2,J,G}.  

We can get a better understanding by identifying the diagrams behind
these four terms.  
The first term $\Delta m_{(+1,0)}$, is the dominant contribution in the
large 
$N_c$ limit.  
It comes from Fig.~2c, where a pion attaches to different quark lines at
the two vertices.  
There is a combinatoric factor $N_c$ at each vertex. As a result the 
sum is proportional to $N_c^2/f^2\sim N_c$.  
Since there is no quark loop in these diagrams, they are completely
unaffected by the introduction of ghosts.  
Hence we come to the conclusion that, {\sl at leading order of $1/N_c$,
chiral one-loop corrections to baryon masses are independent of the
degree of quenching}. 
Fig.~2c contributes at both the fully quenched and the unquenched
limits, 
in sharp constrast to the PQ$\chi$PT for mesons, where the diagrams
appear
in the fully quenched limit always vanish in the unquenched limit, and
{\it
vice versa}.  
In passing, we note that the higher order correction $\Delta m_{(-1,0)}$
also 
comes from the Fig.~2c.  

The term $\Delta m_{(0,+1)}$ comes from Fig.~2a, where the pion attaches
to 
the same quark line at both vertices.  
As a result, the combinatoric factor is $N_c$, and the whole term is of
the 
order $N_c/f^2\sim N_c^0$.  
Fig.~2a has an internal loop, which may be a quark or a ghost.  
As a result, it is sensitive to quenching, hence, proportional to 
$\Delta n$.  
Lastly, $\Delta m_{(-1,-1)}$ is the $\eta'$ loop contribution.  
It is $1/N_c^2$ suppressed with respect to the leading pion loop
correction 
as discussed before, and the $1/\Delta n$ factor comes from the hairpin 
term in the propagator.  
Actually, $\Delta m_{(-1,-1)}$ contains implicit suppression factors as
the hairpin diagrams are OZI, hence, $1/N_c$ suppressed.  
We will study this in detail in section 3.5.  

\subsection{Quenching-Senstivity of Leading Order Chiral Corrections}

In the previous section, we have shown that at leading order of $1/N_c$,
chiral one-loop corrections to baryon masses are independent of $\Delta
n$.  
A casual reader may suggest that this result is a rather trivial
observation.  
Isn't it true that, since internal quark loops are suppressed by $1/N_c$ 
\cite{T}, the effect of quenching must vanish in the large $N_c$ limit
uniformly for any degree of quenching, hence, our result trivially follows?  

The above line of reasoning actually turns out fallacious.  
While it is true that quark loops are suppressed by $1/N_c$, it is
definitely 
not true that the chiral corrections are independent of the degree of
quenching 
at leading order correction of $1/N_c$ expansion. There are numerous 
counter-examples that exhibits large $N_c$ corrections that depend
sensitively
on the degree of quenching.  
As mentioned above, the mass of the $\rho$ meson receives a chiral
correction 
which 
is proportional to $m^3$ in unquenched chiral perturbation theory, but
this 
piece is absent in the quenched theory \cite{boothfalk}, and is in fact 
proportional to $\Delta n$ in the partially quenched theory
\cite{chowrey2}.  
Similarly, the chiral correction to heavy meson mass is non-zero in the 
unquenched theory but vanish in the fully quenched limit 
\cite{booth,sharpezhang}.  

To see flaws in the above argument, note that chiral corrections 
always come from one Goldstone boson loop level.  
For the case of the $\rho$ meson and heavy mesons, all Goldstone boson
loop 
diagrams always contain a quark loop.  
Hence the quark loop enter at leading order of the $1/N_c$ expansion,
and 
hence the chiral corrections are sensitive to the degree of quenching.  
On the other hand, there are Goldstone boson loop diagrams contributing
the 
baryon mass correction which does not contain any internal quark loop 
(Fig.~2c), which is also the leading order contribution in large $N_c$.  
As a result, the leading order result $\Delta m_{(+1,0)}$ is 
$\Delta n$-independent.  

From the above discussion, it should be clear that 
the question whether the leading order 
(in $1/N_c$) chiral correction of a given matter field is
quenching-sensitive 
is highly non-trivial.  It simply is {\it not\/} true that for the
chiral 
corrections for mesons are always quenching-sensitive while those for
baryons 
are quenching-independent.  
In fact one can show that the chiral corrections of flavor non-singlet
mesons 
and baryons with two or more light quarks are always 
quenching-sensitive~\cite{Chow}.  

Lastly, we will discuss the implication of the quenching-independence of 
$\Delta m_{(+1,0)}$.  
We have shown that the effects of quenching are $1/N_c$ suppressed.  
While we cannot prove that these $1/N_c$ suppressed effects are actually 
negligible (they may come with huge coefficients), it is a well-known
rule of 
thumb in hadron phenomenology that $1/N_c$ corrections are usually small 
in comparison with the leading order result if the latter is
non-vanishing.  
In this case, this rule of thumb provides circumstantial evidence that 
the chiral corrections to baryon masses are small, and gives us more 
confidence in the utility of quenched QCD as an approximation of the 
real world of QCD.  

\subsection{Fully Quenched QCD Limit: $\Delta n = 0$}
As recalled earlier, the partially quenched QCD can interpolate and 
bridge between the two extreme limits, viz. fully quenched and
unquenched 
QCD theories by dialing the analytic parameter $\Delta n$ between zero
and $n$. For example, one would expect that by setting $\Delta n = 0$
the 
previously known Q$\chi$PT results should be recovered.
In particular, strongly infared singular non-analytic corrections should 
reappear.
In this section, we will study the fully quenched limit carefully to
understand how these quenched chiral singularities arise. 

First of all, we note that, in the fully quenched limit, $\Delta n\to
0$, the
contributions $\Delta m_{(+1,0)}$ and $\Delta m_{(-1,0)}$ remain intact
and $\Delta m_{(0,+1)}$ vanishes identically. On the other hand, the
contribution $\Delta m_{(-1,-1)}$ apparently diverges in this limit.  
Note, however, that $\Delta m_{(-1,-1)}$ can be re-expressed as
\begin{equation}
\Delta m_{(-1,-1)} = g^2 \, {\bf J}^2 \cdot {2\over\Delta n} \cdot
\left( \quad \left({1\over 1+\Delta n \cdot A_0} \right) \cdot
{\cal I}_1 \left({m^2 + \Delta n \cdot m_0^2\over 1+\Delta n \cdot A_0} 
\right) -{\cal I}_1 (m^2) \quad \right).  
\end{equation}
As $\Delta n \to 0$, the contribution can be re-expressed as a
derivative
with respect to the interpolating parameter $\Delta n$:
\begin{equation}
\Delta m_{(-1,-1)} =  2 g^2 \, {\bf J}^2 \, {d\over d \Delta n} \,
\left[ \quad {1\over 1+ \Delta n \cdot A_0} \cdot {\cal I}_1 \left( \,
{m^2 + \Delta n \cdot m_0^2 \over 1 + \Delta n \cdot A_0}
\, \right)\quad \right]_{\Delta n = 0}.
\label{-1-1}
\end{equation}
Now, it should become clear how the quenched infrared singularity
arises. 
While the contributions from both the pion pole and the shifted pole
have 
the same functional form as the one denoted as ${\cal I}_1$, 
in the fully quenched limit,
the shifted pole returns back to the unshifted position and gives rise
to a 
derivative contribution, which is {\it not\/} of the form ${\cal I}_1$.
In fact, this yields the precise statement of Bernard--Golterman's third 
theorem for baryons in the fully quenched limit.
Expanding the derivative in Eq.~(\ref{-1-1}) explicitly, one finds that
\begin{eqnarray}
&& {d\over d \Delta n} \,
\left[ \quad {1\over 1+\Delta n \cdot A_0} \cdot {\cal I}_1 \left(
{m^2 + \Delta n \cdot m_0^2 \over 1 + \Delta n \cdot A_0} \right) \quad 
\right]_{\Delta n = 0}
\nonumber \\
&= & \, {1\over 12\pi f^2} \, \Big({3\over2} \cdot m_0^2 \cdot 
m - {5\over2}\cdot A_0 \cdot m^3 \Big) 
\equiv {\cal I}_2(m^2).
\end{eqnarray}
Note the appearance of the term linear in $m$. This is the anticipated 
contribution of non-analytic infared singularity in quenched QCD.
To summarize, we conclude that 
the mass correction in the fully quenched limit is
reproduced correctly and is given by 
\begin{equation}
\Delta m = g^2 \left( {3\over8} N_c^2 + {1\over2} {\bf J}^2 \right)
\cdot {\cal I}_1(m^2) + 2 {\bf J}^2 \cdot {\cal I}_2(m^2).
\end{equation}

\subsection{Large-$N_c$ Suppression of $\Delta m_{(-1,-1)}$}
As mentioned before, the hairpin diagram is OZI suppressed.  
As a result, the hairpin parameters, $A_0$ and $m_0^2$, are of order 
$1/N_c$.  
In light of this piece of new information, we will reanalyze the
$1/N_c$ and $\Delta n$ dependence of $\Delta m_{(-1,-1)}$, the term
which
originates from the hairpin propagators.  

To make the $N_c$ dependences explicit, rescale the hairpin
parameters as 
\begin{equation}
m_0^2 = \bar m_0^2/N_c, \qquad A_0 = \bar A_0/N_c, 
\end{equation}
where $\bar m_0^2$ and $\bar A_0$ have smooth non-trivial large $N_c$
limit. 
Then the mass correction becomes 
\begin{equation}
\Delta m_{(-1,-1)} = g^2 {\bf J}^2 \cdot 
\left( {2\over\Delta n} \right) \cdot \Big[ \left( {1\over 1+ (\Delta
n/N_c) 
\bar A_0} \right) \cdot {\cal I}_1 \left(
{m^2 + (\Delta n/N_c) \bar m_0^2\over 1+(\Delta n/N_c) \bar A_0}
\right)-{\cal I}_1(m^2) \, \Big]. 
\label{o} 
\end{equation}
This can be expand in a Taylor series of $\Delta n/N_c$, 
\begin{eqnarray}
\Delta m_{(-1,-1)} &&\!\!\!\!\!\!\!\!\!\!=\! 2 g^2 {{\bf J}^2
\over\Delta n} \!
\sum_{k=1}^\infty \!{(\Delta n / N_c)^k \over k!} 
{d^k \over d(\Delta n / N_c)^k} \!\left[ {1\over 1+ (\Delta n / N_c)\bar
A_0} 
{\cal I}_1 \!\left( {m^2 + (\Delta n / N_c)
\bar m_0^2 \over 1 + (\Delta n / N_c) \bar A_0} \right) 
\right]_{(\Delta n/ N_c)=0} \nonumber\\
&&\!\!\!\!\!\!\!\!\!\!=\! {2g^2 {\bf J}^2 \over N_c} \left[ 
\bar{\cal I}_2(m^2) +{\Delta n\over N_c} \bar {\cal I}_3(m^2) +
\cdots\right] ,
\label{n}
\end{eqnarray}
where $\bar{\cal I}_{2,3}$ are defined as 
\begin{eqnarray}
\bar{\cal I}_2(m^2) &\equiv & {d\over d (\Delta n / N_c)}
\left[ \, {1\over 1+ (\Delta n / N_c)\bar A_0} \cdot {\cal I}_1 \left(
{m^2 +  (\Delta n / N_c)\bar m^2_0 \over 1 + (\Delta n / N_c)\bar
A_0}\right) 
\right]_{(\Delta n / N_c)= 0} \nonumber \\
&= & \, {1\over 12\pi f^2} \, \left( \, {3\over2} \cdot \bar m^2_0 \cdot 
m - {5\over2}\cdot \bar A_0 \cdot m^3 \, \right),
\end{eqnarray}
\begin{eqnarray}
\bar{\cal I}_3(m^2) &\equiv & {d^2\over d(\Delta n / N_c) ^2}
\left[ {1\over 1+  (\Delta n / N_c)\bar A_0} \cdot {\cal I}_1 \left(
{m^2 +  (\Delta n / N_c)\bar m^2_0 \over 1 + (\Delta n / N_c)\bar A_0}
\right)
\, \right]_{(\Delta n / N_c) = 0} \nonumber \\
&=& {1 \over 12 \pi f^2} \left({3 \over 4} \cdot \bar m_0^4 \cdot {1
\over m} 
- {15 \over 2} \bar A_0 \cdot m_0^2 \cdot m - {35 \over 4} \cdot 
\bar A_0^2 \cdot m^3 \right)
\end{eqnarray}

{}From Eq.~(\ref{n}) one can see clearly that the leading $\eta'$
correction 
is of order $g^2/f^2 N_c \sim 1/N_c^2$, viz.~suppressed by one {\sl
higher }
order
in $1/N_c$ than we have naively expected. Moreover, all the higher order 
corrections are in positive powers of $\Delta n$, which vanish
identically in 
the fully quenched limit. The real significance of this result is that, 
in the leading order of an $1/N_c$ expansion which keeps only the first
term 
of Eq.~(\ref{n}), 
$\Delta m_{(-1,-1)}$ is of the form of $\bar {\cal I}_2(m_\pi^2)$, a
form 
which leads to new chiral divergences such as $m_q^{1/2}$, even when the 
theory is far from the fully quenched limit $\Delta n\to 0$!  
In other word, through taking the large $N_c$ limit, one can bypass the 
Bernard--Golterman's third theorem, and have quenched chiral
singularities 
even if the theory is only partially quenched.  
Note that our result is {\it not\/} in contradiction with the 
Bernard--Golterman's third theorem: the analysis of Ref.~\cite{bg2}  
is for $N_c=3$.  
In fact, the third theorem holds for any fixed value of $N_c$ but fails 
only when $N_c$ is taken strictly to infinity.  
It is important to understand that these quenched infared singularities 
appear when the shifted pole coincide with the pion pole.  
Since the shift of the pole is proportional to $\Delta n/N_c$, these 
new chiral divergences appear both at the fully quenched limit and also 
at large (but not infinite) $N_c$ limit.  

We note in passing that the higher order terms in the Taylor expansion 
Eq.~(\ref{n}) are more singular in the chiral limit.  
In the chiral limit, while ${\cal I}_1$ itself goes like $m^3$, 
$\bar{\cal I}_2$ and 
$\bar{\cal I}_3$ behave like $m$ and $1/m$ respectively.
It is easy to verify that each term will be more singular by a factor of 
$1/m^2$.  
If we attempt to take the chiral limit in expansion Eq.~(\ref{n}) before
taking the large $N_c$ limit, all the terms (except $\bar{\cal I}_2$) 
will diverge, and the series will be a sum of infinitely many
infinities.  
However, all these divergences are not physical, as the whole series can
be exactly resummed into the closed form expression Eq.~(\ref{o}).  
The infinities just assert that the true expansion parameter in the
Taylor series is $(\Delta n/2N_c)(\bar m_0^2/m^2)$, which diverges in
the 
chiral limit.  
As a result, the Taylor expansion has behaved badly.  
In other words, the chiral divergences of $\bar{\cal I}_k (k>2)$ are
merely an artifact of applying an expansion outside its radius of 
convergence (zero in this case), while all physically relevant
quantities can be obtained from Eq,~(\ref{o}), which has indeed a 
well-defined chiral limit.  

This above discussion poses an important implication on whether we can
really 
see the $m \sim m_q^{1/2}$ quenched chiral singularities from lattice
data.  
The $\bar {\cal I}_2$ term dominates the Taylor expansion only when 
$m^2 \geq (\Delta n/2N_c)(\bar m_0^2) = \Delta n m_0^2/2$, and 
$\Delta m_{(-1,-1)}$ is linear in $m$.  
With small pion mass $m^2 \leq \Delta n m_0^2/2$, however, 
$\Delta m_{(-1,-1)}$ will deviate from linearity in $m$, as the higher 
$\bar {\cal I}$ are no longer negligible.  
On the other hand, $\Delta m_{(-1,-1)}$ is usually dominated by 
$\Delta m_{(+1,0)}$ except at very low pion masses, {\it i.e.}, $m^2
\leq 8 
{\bf J}^2 m_0^2 /N_c^2$.  
So we can observe $\Delta m$ with a linear dependence on $m$ only over 
the window 
\begin{equation}
\Delta n m_0^2/2 \leq m^2 \leq 8 {\bf J}^2 m_0^2 /N_c^2, 
\end{equation}
and such a window exists only if 
\begin{equation}
\Delta n \leq 16 {\bf J}^2/N_c^2.  
\label{win}
\end{equation}
This condition is independent of the value of the parameter $m_0^2$, and 
is marginally satisfied for $\Delta n=1$, $N_c=3$ and ${\bf J} =
{1\over2}$.  
While this is only an order of magnitude estimate, it suggests that it 
would be difficult to see the linear dependence; we expect that 
of the Bernard--Golterman's third theorem is bypassed 
only when $N_c\to\infty$, but in 
that limit the right hand side of Eq.~(\ref{win}) goes to zero, and the 
window just disappears.  
Physically it means that, when $N_c\to\infty$, the linear term 
is overwhelmed by the leading order term $\Delta m_{(+1,0)}$, which is
larger 
than the linear term by a factor of $N_c^3$.  
We conclude by restating that, while the Bernard--Golterman's third
theorem 
is bypassed in the large $N_c$ limit, it would probably be difficult to 
observe in real lattice data.  

\subsection{Unquenched QCD Limit -- $k=0$}
Let us now consider the conventional QCD with no ghost quarks by taking
the limit $k \to 0$, viz. $\Delta n = n$. We continue working on theory 
with two flavors only. Thus, Eq.~(\ref{m}) becomes
\begin{equation}
\Delta m = g^2 \left[ \quad \left( {3\over8} N_c^2 + {3\over 4} n N_c
+( {1\over2} - {2 \over n}) {\bf J}^2 \right) \cdot {\cal I}_1(m^2) + 
{\bf J}^2 \cdot {2/n \over 1+ n \cdot A_0} \cdot {\cal I}_1 \left( {m^2
+ n\cdot m_0^2 \over 1+ n \cdot A_0} \right)\quad \right].
\end{equation}
Note that the term proportional to ${\cal I}_1(m^2)$ is exactly the
loop correction with one-pion exchange in the standard $n=2$ $\chi$PT.
The interpretation of the shifted pole is in the second term is that 
it is the reinstatement of the contribution of $\eta'$ meson. The shift
of the pole just reflects the fact that, in the real world, the $\eta'$
mass is shifted from the pion mass due to the nonperturbative
resummation of necklace diagrams.

To see this clearer, let us rearrange Eq.~(\ref{m}) in the following
way:
\begin{equation}
\Delta m = g^2 \cdot \left( \quad {3\over8}N_c^2 + {3\over 4}n N_c
+({1\over2}-{2 \over n}) {\bf J}^2 \quad \right) \cdot {\cal I}_1(m^2)
+ g_{\eta'BB}^2 \cdot {\cal I}_1(m_{\eta'}^2),
\label{twoterm}
\end{equation}
in which
\begin{equation}
g_{\eta'BB}^2 \equiv {2 \,{\bf J}^2 \,g^2 / Zn}, \quad
m_{\eta'}^2 \equiv (m^2 + n \cdot m_0^2)/Z, \qquad
{\rm where} \qquad Z \equiv (1 + n \cdot A_0).
\end{equation}
These represent coupling, mass and wave function renormalizations of the
$\eta'$ meson.
Note that the expression for $m_{\eta'}$ and $Z$ agrees with the
standard 
$\chi$PT results in Eq.~(\ref{ep}).  

The second term in Eq.~(\ref{twoterm})
represents the contribution of $\eta'$-meson to the chiral one-loop
correction to the baryon mass spectrum. 
We thus conclude that the unquenched limit $k=0$
does {\sl not} reduce to the $\chi$PT in which only octets of Goldstone
mesons are retained. Instead we have reproduced the $\chi$PT 
{\it without integrating out the $\eta'$ meson}.
This alerts us that we have to be careful when applying the 
Bernard-Golterman's
first theorem, which states that {\sl in the subsector where all valence
quarks are unquenched, the {\rm SU($n|k$)} theory is completely
equivalent to a
normal, completely unquenched {\rm SU($n-k$)} theory.}
We found that this is indeed true, but unlike the standard $\chi$PT,
this ``completely unquenched SU($n-k$) theory'' contains an $\eta'$
meson which may (and does) contribute to chiral loop corrections.  
As a result, naive comparisons between results of (P)Q$\chi$PT and 
their counterparts in standard $\chi$PT may be problematic or
misleading.  
One should instead compare results of (P)Q$\chi$PT with their
counterparts 
in $\chi$PT with $\eta'$ loops.  

In passing we observe that it is impossible to compare 
results of standard $\chi$PT with their counterparts in (P)Q$\chi$PT
with 
the $\eta'$ integrated out.  
In (P)Q$\chi$PT, the $\eta'$ propagator always has a single pole at
$m^2$ 
except for the unquenched limit $k=0$.  
Consequently, there is no $\eta'$--
impossible to integrate out the $\eta'$ meson.  
So the scheme suggested above, namely comparing results of (P)Q$\chi$PT 
with their counterparts in $\chi$PT with $\eta'$ loops, is the only 
option which is theoretically well-justified.  

\section{Discussion}

\subsection{$1/N_c$ Corrections}
So far, we have kept only contributions that are leading order in
$1/N_c$ expansion.  In this section we will discuss several sources of
higher-order $1/N_c$ corrections we have ignored so far.  
First of all, there are corrections to the axial current couplings to 
baryons through multi-quark operator.  
It has been shown \cite{J} that the leading order correction to pion 
coupling is through the two-quark operator 
\begin{equation}
{1\over N_c} \, {g'\over f} \, \left(\partial^i \pi^a \right) 
\sum_{k,\ell=1}^{N_c} 
\left(q_k^\dag ({\bf S}^i \otimes {\bf 1}) q_k \right) 
\left(q_\ell^\dag ({\bf 1} \otimes {\bf T}^a) q_\ell \right), 
\end{equation}
where $g'$ is an undetermined coupling constant of order ${\cal
O}(N_c^0)$.  
This will lead to new single-pion loop corrections: a $gg'/f^2$ term at
the 
order of ${\cal O}(N_c^{-1})$, and a $g'^2/f^2$ term at the order of 
${\cal O}(N_c^{-3})$.  
Since these terms are sub-leading with respect to the leading pion
contribution
$\sim {\cal O}(N_c)$, they can be safely ignored in the large $N_c$
limit.  
What cannot be ignored is the $\eta'$ counterpart.  
By planar symmetry, $\eta'$ can also couple to a baryon through the
following two-quark operator:  
\begin{equation}
{1\over N_c} \, {g'\over f} \, \left( \partial^i \eta' \right) \, 
\sum_{k,l=1}^{N_c} 
\left(q_k^\dag ({\bf S}^i \otimes {\bf 1}) q_k \right) \left(
q_l^\dag ({\bf 1} \otimes {\bf 1})  q_l \right) = 
{g'\over f} \, \partial^i \eta' \sum_{k,l=1}^{N_c} 
\left(q_k^\dag ({\bf S}^i \otimes {\bf 1}) q_k \right), 
\end{equation}
which is of the same form as the leading order $\eta'$ coupling.  
In a consistent $1/N_c$ expansion, one has to keep the effect of $g'$,
which contribute in leading order (${\cal O}(N_c^{-1})$) to the quenched 
infared divergences $\Delta m_{(-1,-1)}$\footnote{ The coupling $g'$
appears also in the other contributions to $\Delta m$, but the effects
are subleading.} :
\begin{equation} 
\Delta m_{(-1,-1)} = (g+g')^2 {\bf J}^2 {2\over\Delta n} 
({\cal I}_1(M^2) - {\cal I}_1(m^2)) \sim {\cal O}\left( N_c^{-1} 
\Delta n^{-1}
\right).  
\end{equation}
This does not change the form of the new chiral singularities, but its 
coefficient is no longer related to the standard $\chi$PT correction 
($\Delta m_{(1,0)}$).  

Another possible source of correction is the inclusion of the hairpin 
coupling $h$, which is again of order $1/N_c$.  
We have not calculated $\Delta m$ in the presence of a non-zero hairpin 
coupling, which should be straightforward but adds significantly more
complications.  
However, the functional form of $\Delta m$ is not affected by the
inclusion of $h$, as the contributing Feynman diagrams are still of the 
same form as those shown in Fig.~1. Hence, we expect that 
only the overall factor $g^2$ will be modified: 
\begin{equation}
g^2 \to (g + x h)^2 .  
\end{equation}
Unfortunately, the variable $x$ may be different for different
contributions 
for $\Delta m$, {\it i.e.}, $x$ for $\Delta m_{(+1,0)}$ may be different 
from that of $\Delta m_{(-1,-1)}$.  
However, since $h$ is $1/N_c$ suppressed with respect to $g$, the change
will 
only be higher order effects in a $1/N_c$ expansion.  
In any case, our analysis of the fully quenched and the unquenched limit 
depends only on the functional form taken by $\Delta m$.  
The coupling constants are irrelevant for this analysis, and we expect
the reported results remain valid even with a non-zero hairpin coupling, 
even though the details have to be checked.  

\subsection{Conclusion}
In this paper, we have studied baryons in partially quenched large-$N_c$
QCD with particular emphasis to the interpolation between fully
unquenched
and fully quenched limits. 
In the large $N_c$ limit of partially quenched QCD, we have calculated 
$\Delta m$, the chiral one-loop correction to baryon masses. 
The main results are:  

$\bullet$  The leading order contribution to $\Delta m$ is of order
${\cal O} (N_c)$, in agreement of the large $N_c$ counting rules.  
This contribution does not depend on the value of $\Delta n$, which
survives both in the fully quenched and the unquenched limits.  
In other words, the leading term of $\Delta m$ is independent of 
number of ghost quarks introduced.  
This provides circumstantial evidence that quenching correction to 
baryon masses is small.  

$\bullet$  For any fixed $N_c$ our expression for $\Delta m$ does
satisfy the Bernard--Golterman's third theorem and has no quenched
infared 
singularities except $\Delta n\to0$ limit.
In the large $N_c$ limit, however, the third theorem is bypassed. We 
indeed find appearance of the quenched chiral corrections, which are 
unfortunately of order ${\cal O}(N_c^{-2})$ and $1/N_c^3$ suppressed
with 
respect to the leading order contribution to $\Delta m$ and hence making 
them difficult to be observed.  

$\bullet$  In the unquenched limit, $\chi$PT results are reproduced with 
$\eta'$ loop contributions.  

Unfortunately, it is highly non-trivial to relate our result with that 
of Ref.~\cite{labrenzsharpe}, which has studied Q$\chi$PT for baryons
with
$n=k=3$ and $N_c=3$.  
The reason is that it turns out that the spin-flavor structure of large 
$N_c$ baryons with $n=3$ are much more complicated than that in $n=2$.  
In particular, the large $N_c$ limit of the ratios between the SU(3)
coupling 
constants $\cal F$, $\cal D$, $\cal C$ and $\cal H$ in 
Ref.~\cite{labrenzsharpe} are ambiguous \cite{DJM1}.  
The study and clarification of these issues are straightforward, but is 
beyond the scope of the present paper.  
Being that the motivation and the focus of our study in the present
paper
are of  more theoretical issues than phenomenological ones, 
detailed comparison of our results with
those of Ref.~\cite{labrenzsharpe} will be relegated in a separate paper
elsewhere.

We will end the paper by discussing possible relevance of our studies.  
Our motivations are mainly theoretical and we have tried to disentangle 
the interplays between chiral dynamics, large $N_c$ expansion and 
quenching.  
On the other hand, while most of the lattice QCD simulation are
performed 
with $N_c=3$, there are also simulations with arbitrary $N_c$, most
notably 
by Teper \cite{Teper}.  
So far, these studies have focused on pure Yang-Mills theory and have
studied 
string tension and glueball masses. However, new data on large $N_c$ QCD
should
become available in the future.  
More phenomenologically, there has been a lot of investigation 
of baryon properties in the large $N_c$ limit
\cite{DJM1,CGO,LM,DJM2,J,G}, 
with many new interesting results.  
For example, the $\Delta$-N splitting is expected to be of order
$1/N_c$.  
While these results are successful in organizing the baryon properties, 
one would like to directly verify these predictions with the $\Delta$-N 
splitting for different values of $N_c$.  
This, however, is only possible with lattice simulations. 
Our studies should be relevant if these simulations are really
undertaken 
in the foreseeable future.  
In conclusion, we are using the $1/N_c$ expansion as a guiding principle
to 
help identifying and organizing spectrums and chiral corrections when
one 
works on quenched lattice world, hoping that $1/N_c$ expansion is as 
successful in hadrons phenomenology on the lattice as it has been in the 
real world.

\end{document}